# Electrical Characteristics of Superconducting-Ferromagnetic Transistors

Ivan P. Nevirkovets, Oleksandr Chernyashevskyy, Georgy V. Prokopenko, Oleg A. Mukhanov, *Fellow, IEEE,* and John B. Ketterson

*Abstract*—We report experimental results on characteristics of SFIFS junctions and multi-terminal SFIFSIS devices (where S, I, and F denote a superconductor (Nb), an insulator ($AlO_x$), and a ferromagnetic material (Ni), respectively). The SFIFS junctions serve as injectors in the SFIFSIS devices which have transistor-like properties; for this reason we call them Superconducting-Ferromagnetic Transistors (SFTs). We have found the F (Ni) thickness at which the SFIFS current-voltage characteristic (CVC) becomes linear. Furthermore, we investigated the DC and AC characteristics of SFTs of two types: ordinary devices with a single acceptor (SIS) junction, and devices with a double acceptor. In the first case, we focused on studying the influence of the injection current through the SFIFS junction on the maximum Josephson current of the SIS acceptor. For devices of the second type, we studied voltage amplification properties when the operating point was chosen in the sub-gap region of the acceptor CVC. By applying an AC signal (in the kHz range) while biasing the injector (SFIFS) junction with a constant DC current, we observed a voltage gain above 25 on the double acceptor. In the reverse transmission experiment, we applied DC current and an AC modulation to the acceptor junction and, within the accuracy of the experiment, observed no response on the injector junction, which implies an excellent input-output isolation in our SFIFSIS devices. The experiments indicate that, after optimization of the device parameters, they can be used as input/output isolators and amplifiers for memory, digital, and RF applications.

## I. Introduction

IN SPITE of considerable progress in the development of ultra-sensitive devices based on superconductivity, the enormous potential of the superconducting state has yet to be fully exploited. In particular, there is a major need for a superconducting three-terminal device with both gain and input/output isolation. Lack of such a device is a severe barrier limiting radio-frequency, and ultra-fast, low-power digital superconducting electronics applications. Numerous designs for a superconducting transistor have been proposed, but they all have drawbacks preventing their implementation in real circuits [1]-[10]. Materials and structures where the competing phenomena of ferromagnetism and superconductivity coexist have the potential to impact this problem.

Earlier we reported on SIFSIS [11], [12] and SFIFSIS [13], [14] multi-terminal devices with transistor-like properties (where S, I, and F denote a superconductor (Nb), an insulator ($AlO_x$), and a ferromagnetic material (Ni), respectively). It is our contention that such devices will enable voltage, current, and power amplification, while at the same time having good input-output isolation. In particular, the devices may be applied to amplify the low-voltage output of the Single Flux Quantum (SFQ) circuits to the level of about 2 mV suitable for further processing by room-temperature electronics.

Here we present experimental results on characteristics of SFIFS junctions and multi-terminal SFIFSIS devices which involve SFIFS injectors and SIS acceptors (both single and double).

## II. Properties of SFIFS junctions

Since the SFIFS junctions serve as injectors in our SFT devices, it is worthwhile to investigate their properties in a more detail. Specifically, it is interesting to study Josephson tunneling in such junctions in dependence of the F layer thickness, and determine the critical thickness when the current-voltage characteristic (CVC) becomes linear. For this purpose, we have fabricated and characterized three types of Nb/Ni/Al/$AlO_x$/Al/Ni/Nb junctions where the two Ni layers were nominally identical and had thicknesses, $d_{Ni}$, of 0.7, 1.3, and 2.0 nm for each type, respectively.

First, we measured magnetic moment vs. applied magnetic field for parts of unprocessed chips with these structures at 10 K (i.e., just above the critical temperature of Nb). The measurements were carried out in a Quantum Design MPMS system. The data for the structures with $d_{Ni}$ = 0.7 nm (red circles) and $d_{Ni}$ = 2.0 nm (black squares) are shown in Fig. 1; the data for the structure with $d_{Ni}$ = 1.3 nm are similar to those for the structure with $d_{Ni}$ = 0.7 nm. One can see that clear hysteretic behavior appears for the structure with $d_{Ni}$ = 2.0 nm, whereas the hysteresis is not resolved (within the accuracy of the measurement) for the structures with thinner Ni. This means that magnetic ordering in the Ni layers begins when the layer thickness reaches about 2 nm.

We observe a correlation between the above property and the CVC of the respective junctions (characterized at 4.0 K);

This work was supported in part by IARPA under ARO contract #W911NF-09-C-0036.

I. P. Nevirkovets and O. Chernyashevskyy are with the Department of Physics and Astronomy, Northwestern University, Evanston, IL 60208 USA; e-mail: i-nevirkovets@northwestern.edu; o-chernyashevskyy@northwestern.edu.

G. V. Prokopenko and O. A. Mukhanov are with HYPRES, Inc., Elmsford, NY 10523 USA (e-mail: georgy@hypres.com; mukhanov@hypres.com).

J. B. Ketterson is with the Department of Physics and Astronomy, and Department of Electrical Engineering & Computer Science, Northwestern University, Evanston, IL 60208 USA; e-mail: j-ketterson@northwestern.edu.

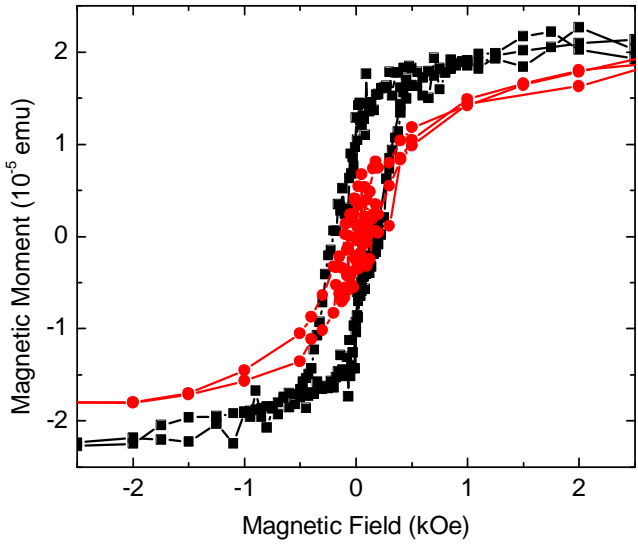

Fig. 1. Magnetic moment as a function of applied field for two Nb/Ni/Al/AlO$_x$/Al/Ni/Nb structures at 10 K. Red circles: $d_{Ni}$ = 0.7 nm; black squares: $d_{Ni}$ = 2.0 nm.

see Fig. 2. (The junction area in all cases was 10 μm × 10 μm). Specifically, a significant Josephson current is observed for $d_{Ni}$ = 0.7 nm, much lower for $d_{Ni}$ = 1.3 nm, and no Josephson current for $d_{Ni}$ = 2.0 nm. In the latter case, the CVC is linear, meaning that the total thickness of two Ni layers (about 4 nm) is larger than the superconducting coherence length in Ni deposited under the conditions of this experiment.

With this knowledge, we can choose appropriate Ni thickness in order to obtain linear CVC of the injector junction in our SFT devices.

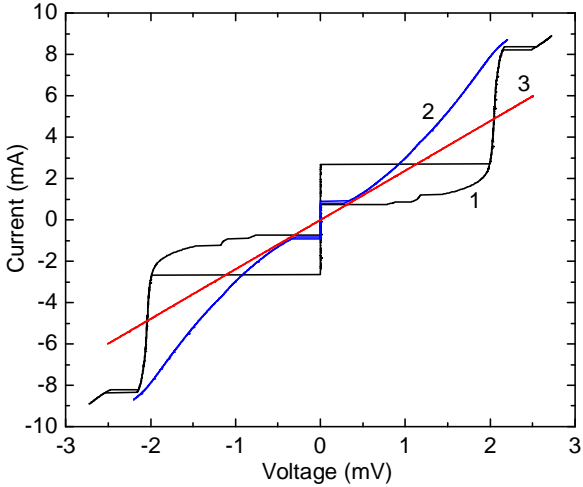

Fig. 2. Current-voltage characteristics (CVCs) of three Nb/Ni/Al/AlO$_x$/Al/Ni/Nb junctions taken at 4.0 K for $d_{Ni}$ = 0.7 nm (black curve 1), $d_{Ni}$ = 1.3 nm (blue curve 2), and $d_{Ni}$ = 2.0 nm (red curve 3). In the latter case, the CVC is linear and shows no Josephson current.

### III. PROPERTIES OF SFIFSIS DEVICES

Next we consider double-barrier SFIFSIS devices (SFTs). The devices were fabricated from Nb/Ni/Al/AlO$_x$/Ni/Nb/Al/AlO$_x$/Nb multilayers deposited on either sapphire or oxidized Si substrates and tested at 4.2 K.

Fig. 3 shows schematic cross-sectional view and biasing for the single-acceptor (a) and double-acceptor (b) SFT device.

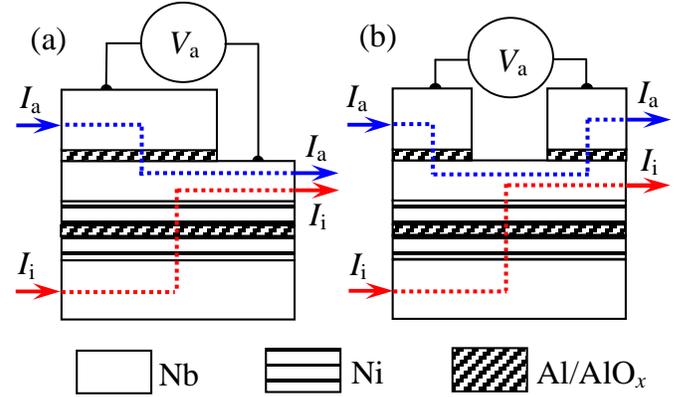

Fig. 3. Schematic cross-sectional view and biasing for the single-acceptor (a) and double-acceptor (b) SFT device.

Here we report the data for three devices, D$_1$-D$_3$. Some parameters for these devices are summarized in Table I. Below we consider characteristics of these devices in a more detail.

TABLE I
SFT DEVICE PARAMETERS

| Device No. | Substrate | No. of acceptors | Injector size (μm × μm) | Acceptor size (μm × μm) | $d_{Nb2}$ (nm) | $d_{Ni}$ (nm) for each of 2 layers | Injector $R_T$ (Ω×cm$^2$) | Acceptor $R_T$ (Ω×cm$^2$) |
|---|---|---|---|---|---|---|---|---|
| D$_1$ | Al$_2$O$_3$ | 1 | 5 × 7.5 | 5 × 5 | 30 | 2 | 3.1 × 10$^{-7}$ | 3.3 × 10$^{-7}$ |
| D$_2$ | Si/SiO$_2$ | 1 | 10 × 12 | 8 × 10 | 45 | 2 | 1.3 × 10$^{-6}$ | 5.5 × 10$^{-7}$ |
| D$_3$ | Al$_2$O$_3$ | 2 | 10 × 12 | 4 × 8 | 35 | 2 | 3.9 × 10$^{-7}$ | 2.5 × 10$^{-7}$ |

#### A. Modulation of Josephson Current in the Single-Acceptor Devices

For some applications, it is important to be able to control Josephson current. These applications include random access memory (RAM), in which an SFT can act as a sell selector similarly to a semiconductor transistor performing this function in room-temperature magnetic RAM (MRAM) [15]-[16]. In cryogenic MRAM, our SFT device can provide this capability and can be integrated with magnetic storage elements being explored today [17]-[23].

Quasiparticle injection from SFIFS junction suppresses the energy gap in the middle (Nb$_2$) electrode, thus resulting in suppression of the Josephson current in the SIS (acceptor) junction. We demonstrate this using the single-acceptor SFTs D$_1$ and D$_2$. Fig. 4 shows CVC of the acceptor (black curve 1) and injector (blue curve 2) junctions for device D$_2$. Red curve 3 is initial portion of the acceptor CVC recorded in an applied magnetic field corresponding to the second minimum of the $I_c$ vs. $H$ dependence (where $I_c$ is the critical Josephson current). This curve displays the gap difference feature, which allows us to determine the superconducting energy gaps of the middle Nb$_2$ and the top Nb$_3$ electrodes to be 0.86 meV and 1.22 meV,

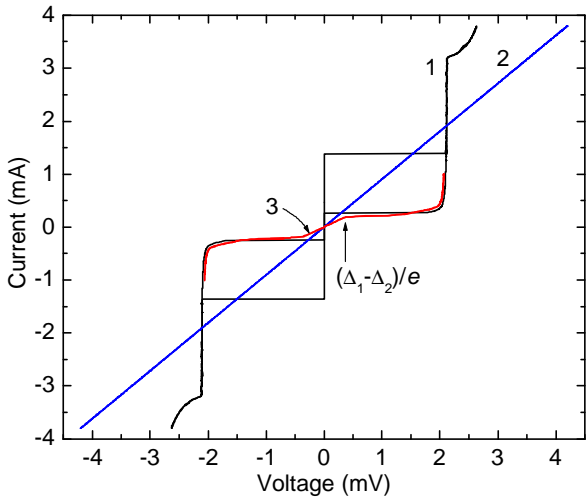

Fig. 4. Current-voltage characteristics (CVCs) for SFT device $D_2$ at 4.2 K. Black curve 1 is CVC of the acceptor SIS junction, blue curve 2 is CVC of the injector SFIFS junction, and red curve 3 is an initial portion of the acceptor CVC in an applied magnetic field corresponding to the 2$^{nd}$ minimum of the $I_c$ vs. $H$ dependence. Gap difference feature is seen in the latter curve.

respectively.

We measured the $I_c$ vs. $H$ dependence for the SIS (i.e., $Nb_2/Al/AlO_x/Nb_3$) acceptor junction of device $D_2$ at different levels of current through the injector junction SFIFS. These data are shown in Fig. 5. Curves from top to bottom are for the injection current, $I_i$, from 0 to 4 mA applied with the 0.4 mA increment. Regular shape of the $I_c$ vs. $H$ dependence is preserved up to high injection current; at $I_i$ = 4.0 mA the dependence is distorted, which may be due to several reasons: (1) trapping the magnetic flux; (2) development of an inhomogeneous gap state under quasiparticle injection on the scale of diffusion length [24], or (3) transition into the π-state under influence of spin injection [25]. The latter case is most interesting from physical point of view. Further experiments are needed to establish the cause of such distortion.

In Fig. 6, we plotted maximum Josephson current as a function of the injection current level for devices $D_1$ and $D_2$.

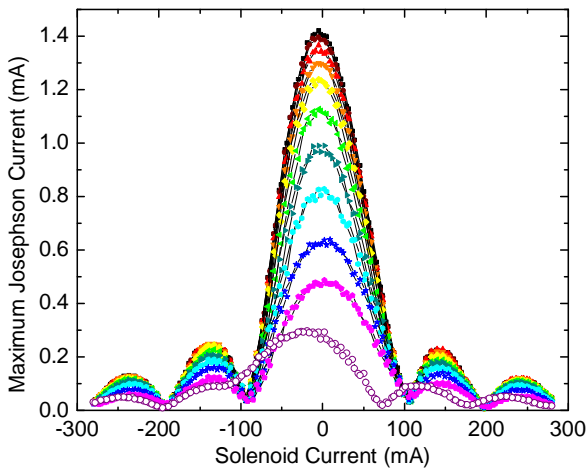

Fig. 5. $I_c$ vs. $H$ dependence for the acceptor junction of the device $D_2$ at different levels of the injection current. Curves from top to bottom are for the injection current from 0 to 4 mA applied with the 0.4 mA increment.

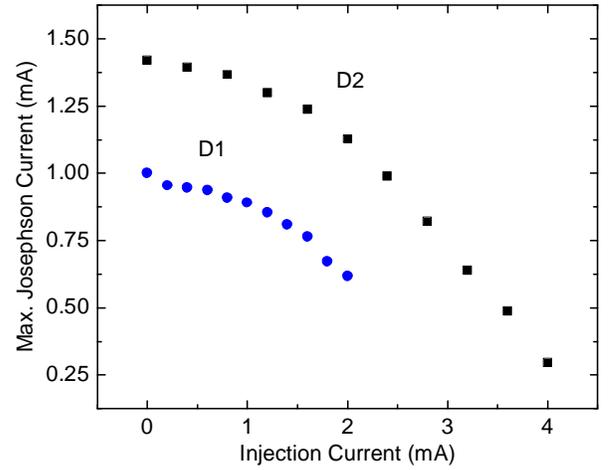

Fig. 6. Maximum Josephson current of the acceptor junction vs. the current through the injector junction for devices $D_1$ and $D_2$.

The data demonstrate possibility to modulate Josephson current by quasiparticle injection in SFT devices. Optimization of the devices is needed in order to achieve more efficient modulation.

B. *Voltage Amplification*

Next we consider voltage amplification in the double-acceptor SFTs exemplified by device 3. The experiment,

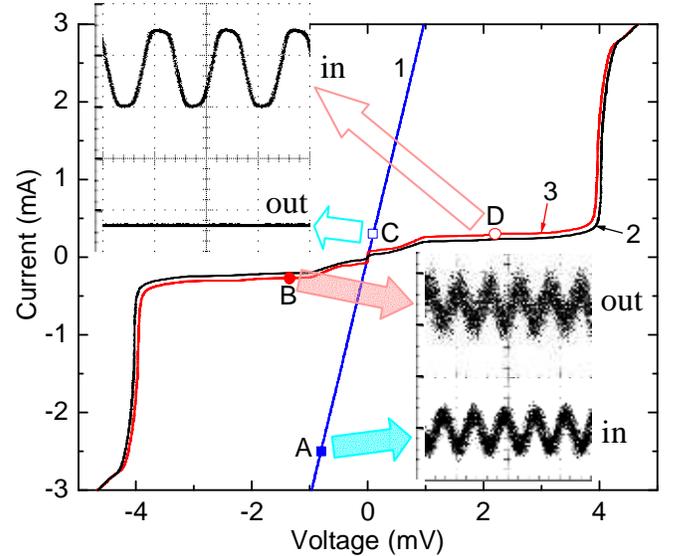

Fig. 7. Illustration of the voltage amplification experiment on device $D_3$. Curve 1 is CVC of the injector; curve 2 is unperturbed CVC of the double acceptor in an applied magnetic field of 250 Oe; curve 3 is the same CVC but under influence of the injection current corresponding to the DC bias point A. If a small AC signal is applied in point A (bottom signal in the lower inset) then one obtains an output signal (shown on top in the lower inset) in the operation point B of the acceptor CVC. The vertical (voltage) scale is 20 μV per division for the input signal, and 500 μV per division for the output signal. (Horizontal scale is 5 ms/division in all cases.) One can infer the voltage gain above 25. In the reverse transmission experiment, the input signal (top curve in the upper inset) was fed at the DC bias point D of the acceptor CVC, and the output signal (thick line in the upper inset) was acquired at the operation point C of the injector CVC. The voltage scale is 500 μV per division for the input signal, and 5 μV per division for the output signal. This experiment indicates very good input/output isolation.

illustrated in Fig. 7, was carried out at 4.2 K. In Fig. 7, Curve 1 is CVC of the injector; curve 2 is CVC of the double acceptor recorded at zero injection current in an applied magnetic field of 250 Oe; curve 3 is the same CVC but under influence of the injection current corresponding to the DC bias point A in curve 1. If, in addition to DC bias current, a small AC signal is applied in point A (an image of the oscilloscope screen displaying this signal is shown on bottom in the lower inset) then one obtains an output signal (shown on top in the lower inset) in the operation point B of the double-acceptor CVC. The vertical (voltage) scale is 20 µV per division for the input signal, and 500 µV per division for the output signal. (Horizontal scale is 5 ms/division in all cases.) The peak-to-peak amplitude of the input signal is 20 µV, whereas the peak-to-peak amplitude of the output signal is about 600 µV; therefore, the voltage gain is about 30. In the reverse transmission experiment, the input signal (top curve in the upper inset) was fed at the DC bias point D of the double-acceptor CVC, and the output signal (thick line on bottom in the upper inset) was acquired at the operation point C of the injector CVC. The voltage scale is 500 µV per division for the input signal, and 5 µV per division for the output signal. From this experiment, one can infer very good input/output isolation in our SFT device.

## IV. CONCLUSION

We fabricated and studied characteristics of both two-terminal SFIFS (Nb/Ni/Al/AlO$_x$/Al/Ni/Nb) and multi-terminal SFIFSIS (Nb/Ni/Al/AlO$_x$/Ni/Nb/Al/AlO$_x$/Nb) devices at liquid He temperatures. We observed reduction of the Josephson current in SFIFS junctions to zero when the thickness of the Ni layer is increased to 2 nm, resulting in linear CVC.

The multi-terminal SFIFSIS devices, which we call superconducting-ferromagnetic transistors (SFTs), were studied in two configurations: single-acceptor and double-acceptor. For the single-acceptor devices, we demonstrated experimentally that the Josephson critical current of the SIS acceptor junction can be efficiently controlled by quasiparticle injection from the SFIFS junction.

For the double-acceptor devices, we explored possibility to amplify a sinusoidal voltage signal fed through the injector junction in addition to a DC bias current. The output signal was measured across the SIS double-acceptor junction when biased in the subgap region of its current-voltage characteristic in the presence of a magnetic field applied to suppress the Josephson current. A voltage gain above 25 was obtained. In the reverse-transmission experiment, a sinusoidal signal was applied to the SIS double-acceptor junction DC biased in the subgap region of the CVC, but no output signal was observed across the injector junction. This means perfect input/output isolation in our SFT devices.

At this stage, little data are collected on operation of SFT devices. Clearly, more experimental and theoretical investigation is needed to optimize the device performance. Our initial study indicates that the SFT devices are promising for various applications in superconducting electronics including energy-efficient superconducting computing circuits [26], [27].

## ACKNOWLEDGMENT

Authors acknowledge useful discussions with I. Vernik and encouragement from M. Manheimer and S. Holmes.